\documentclass[a4paper, amsfonts, amssymb, amsmath, reprint, showkeys, nofootinbib, oneside,superscriptaddress]{revtex4-1}
\usepackage[english]{babel}
\usepackage[utf8]{inputenc}
\usepackage{float}
\usepackage[colorinlistoftodos, color=green!40, prependcaption]{todonotes}
\usepackage{amsthm}
\usepackage{mathtools}
\usepackage{physics}
\usepackage{xcolor}
\usepackage{graphicx}
\usepackage[left=23mm,right=13mm,top=35mm,columnsep=15pt]{geometry} 
\usepackage{adjustbox}
\usepackage{placeins}
\usepackage[T1]{fontenc}
\usepackage{lipsum}
\usepackage{csquotes}
\usepackage[pdftex, pdftitle={Article}, pdfauthor={Author}]{hyperref} 
\begin{document}
\title{Simplex2Vec embeddings for community detection in simplicial complexes}

\author{Jacob Charles Wright Billings}
    \affiliation{Institute for Scientific Interchange, Torino, Italy}

\author{Mirko Hu}
    \affiliation{IMT School for Advanced Studies, Lucca, Italy}   

\author{Giulia Lerda}
    \affiliation{Institute for Scientific Interchange, Torino, Italy}
    
\author{Alexey N. Medvedev}
    \affiliation{ICTEAM, Université Catholique de Louvain, Louvain-la-Neuve, Belgium}
    
\author{Francesco Mottes}
    \affiliation{Università di Torino, Physics Department and INFN, Torino, Italy}
 
\author{Adrian Onicas}
    \affiliation{IMT School for Advanced Studies, Lucca, Italy} 
   
\author{Andrea Santoro}
    \affiliation{School of Mathematical Sciences, Queen Mary University of
  London, London E1 4NS, United Kingdom}

\affiliation{The Alan Turing Institute, The British Library, NW1 2DB,
  London, United Kingdom}

\author{Giovanni Petri}
    \email[Correspondence email address: ]{giovanni.petri@isi.it}
    \affiliation{Institute for Scientific Interchange, Torino, Italy}
    \affiliation{ISI Global Science Foundation, NewYork, USA}

\date{\today} 

\begin{abstract}
Topological representations are rapidly becoming a popular way to capture and encode higher-order interactions in complex systems. They have found applications in disciplines as different as cancer genomics, brain function, and computational social science, in representing both descriptive features of data and inference models.  While intense research has focused on the connectivity and homological features of topological representations, surprisingly scarce attention has been given to the investigation of the community structures of simplicial complexes. 
To this end, we adopt recent advances in symbolic embeddings to compute and visualize the community structures of simplicial complexes. 
We first investigate the stability properties of embedding obtained for synthetic simplicial complexes to the presence of higher order interactions. We then focus on complexes arising from social and brain functional data and show how higher order interactions can be leveraged to improve clustering detection and assess the effect of higher order interaction on individual nodes. We conclude delineating limitations and directions for extension of this work.
\end{abstract}


\maketitle
\section{Introduction}
\setlength{\arrayrulewidth}{1mm}
\setlength{\tabcolsep}{18pt}
\renewcommand{\arraystretch}{1.5}

In the last decades, network science has successfully characterized real-world complex systems ~\cite{Barabasi_Albert_2002,Newman_book2010,Latora_Nicosia_Russo_book2017} by studying each system's corresponding network representation. Nodes represent the elements of the system, and links represent pairwise interactions ~\cite{Latora_Review_2006,barrat2004architecture}.  However, many real-world networks including social systems ~\cite{bajardi2015unveiling,carstens2013persistent,bonato2010geometric,schebesch2017topological,sekara2016fundamental}, collaboration networks among scientists~\cite{carstens2013persistent,patania_shape_2017}, joint appearances among actors~\cite{ramasco2004self}, neural activity of the human brain~\cite{giusti2015clique,petri_homological_2014, lord2016insights,bassett2017network,bendich2016persistent,yoo2016topological}, etc., often exhibit interactions that cannot be captured when only considering pairwise relationships. As an example, consider a congress who meets as a single body when in session. Representing this relationship as a set of pairwise meetings would miss the fact that more than two congresspersons must meet together to make a legal quorum. Moreover, several real-world networks display "small-world" topologies, where high concentrations of edges exist among special groups of vertices, and where low concentrations of edges exist between these groups.

Owing to the importance of understanding correlated processes in terms of their function(s) as a network, much attentions has been devoted to the development of new and scalable methods to extract communities and their boundaries based on the structural roles of interconnected nodes ~\cite{fortunato2010community,lancichinetti2011finding,newman2006modularity,newman2013spectral,javed2018community,patania_topological_2017}. Simplicial complexes were introduced as a way to account for higher-order interactions beyond more traditional pairwise descriptors ~\cite{ghrist2008barcodes,carlsson2005persistence,chazal2017introduction,lambiotte2019networks}. Indeed, it has been shown that simplicial representations provide new and satisfactory explanations for many complex dynamics in neuroscience ~\cite{lord2016insights,bassett2017network,bendich2016persistent,yoo2016topological}, social systems \cite{iacopini2019simplicial,skardal} and dynamical systems \cite{bianconi2018topological,petri_icial_2018,maletic2016persistent}. 
Surprisingly, with the exception of few notable examples~\cite{courtney2016generalized,bianconi2015complex,bianconi2017emergent}, scarce attention has been devoted to define community structure of simplicial complexes.

To this end, the present work implements recent advances in symbolic embeddings, i.e. node2vec and word2vec \cite{grover2016node2vec, goldberg2014word2vec}, combined with biased and unbiased random walks to map, detect and visualize the community structure of simplicial complexes.  

\section{Methods}

\subsection{Simplices and simplicial complexes}
The simplest definition of a $k$-dimensional simplex $\sigma$ is combinatorial: a simplex is a set of $k+1$ vertices  $\sigma = [x_0, x_1, \ldots,x_k]$. It is easy to understand how simplices can describe both pairwise and group interactions. An edge is a collection of two vertice $[x_0, x_1]$, while larger sets represent groups, e.g. a 2-simplex is a ``filled'' triangle $[x_0, x_1,x_2]$, while the set of all its edges is $[x_0,x_1], [x_0,x_2],[x_1,x_2]$. Notice that a set of simplices constitutes simplicial complexes in a similar manner to how a set of edges defines a network. 
More formally, a simplicial complex $\mathcal{K}$ on a given set of vertices $V$, with $|V|= N$, is a collection of simplices, with two conditions: (i) if $\sigma \in \mathcal{K}$ then all the possible sub-simplices $\tau \subset \sigma$  constructed from any subsets of $\sigma$ are also contained in $\mathcal{K}$ and (ii) the intersection of any two $\sigma_{i}, \sigma_{j} \in \mathcal{K}$ is empty or a face of both $\sigma_i$ and $\sigma_j$.

There are several ways to build a simplicial complex. A commonly used one is via clique complexes: we start from a network and promote each $k-$clique $\mathcal{C}$ to $(k-1)$ simplex defined by the nodes of $\mathcal{C}$. Then we focus our attention on maximal cliques, since they are automatically promoted to facets, i.e. simplices that are not a face of another simplex of the complex, and a list of facets uniquely defines a simplicial complex. Moreover, each facet contains all its subsimplices  \cite{petri_homological_2014}.
In a clique complex however it is impossible to have an empty triangle ($[x_0,x_1], [x_0,x_2],[x_1,x_2]$, without $[x_0, x_1, x_2]$), since its structure is defined by its 1-dimensional skeleton, the collection of its edges (1-simplices). For this reason, in section \ref{sec:simpconstruction} we consider alternative ways to build simplices.

\subsection{Hasse diagrams and persistent homology}
\begin{figure*}[t]
    \centering
    \includegraphics[width=0.9 \textwidth]{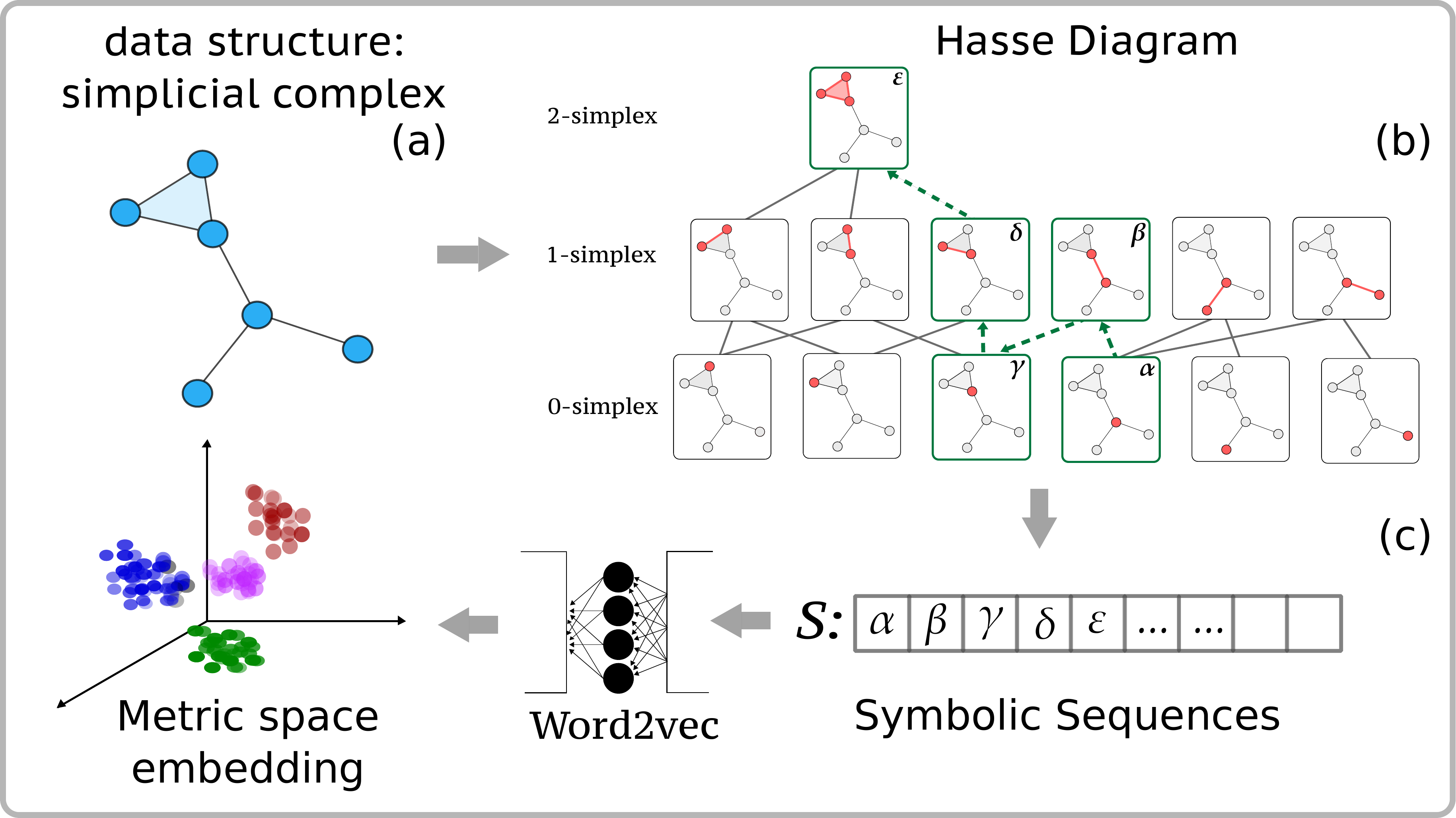}
    \caption{\textbf{Mapping simplicial complexes to metric spaces: Simplex2Vec}. The collection of simplices, representing $d$-dimensional group interactions are glued together in a simplicial complex (a). The simplicial complex is  then represented through a Hasse diagram (b), which is a directed acyclic graph on the partially ordered set of simplices. A random walk on the nodes ($k$-simplex) of the Hasse diagram, graphically represented as a dashed green line in the panel, produces a symbolic sequence which preserve the high-order information on the topological structure of the simplicial complex. The symbolic sequences are then mapped into a metric space using word2vec and finally, an agglomerative clustering method is used to clusterize nodes into different modules.}
    \label{fig:Simplex2Vec}
\end{figure*}

Computational topology and Topological Data Analysis (TDA) are concerned with how to represent and quantify the structure of simplicial complexes~\cite{edelsbrunner2010computational}. 
One of the most common, indeed equivalent, mathematical representations of simplicial complexes is based on the  \textit{Hasse diagram}. 
This is a directed acyclic graph (DAG), where each vertex represents a $k$-simplex of the simplicial complex and there exist an edge connecting two vertices $v_0$ and $v_1$, if $v_0 \subset v_1$ and $dim(v_0) = dim(v_1)-1$, i.e.
they correspond to two simplices of consecutive dimensions (see Figure \ref{fig:Simplex2Vec}b for a graphical representation). 
The Hasse diagram effectively is the backbone of inclusions between simplices of different dimensions. 

\begin{figure}[b]
    \centering
    \includegraphics[width=0.6\columnwidth]{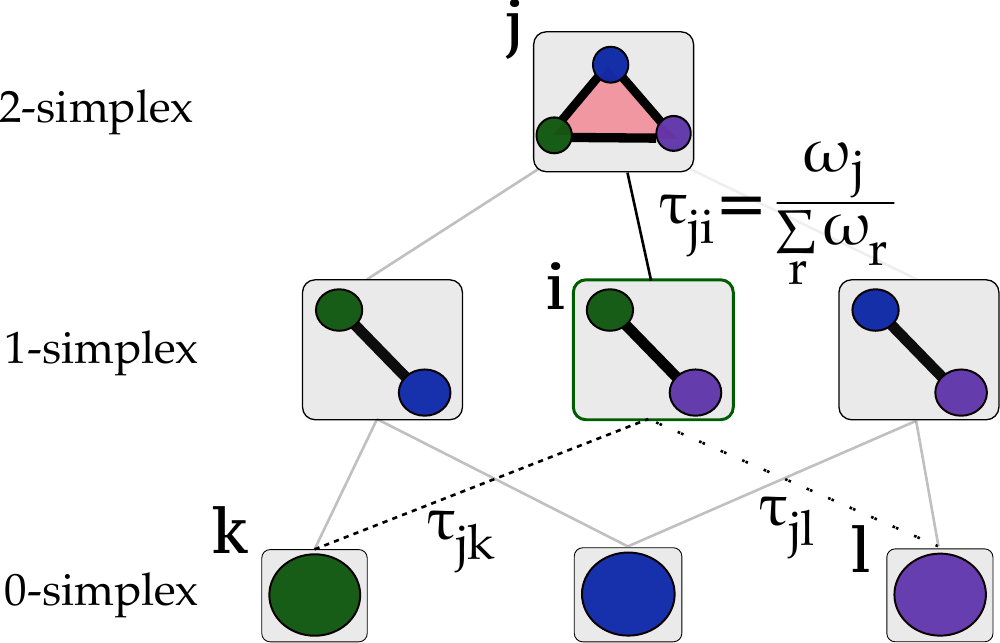}
    \caption{\textbf{Random  walk on simplices}. At each time-step $t$, a walker located on a $k$-simplex $i$ of the Hasse diagram can jump to a simplex $j$ with a certain probability $\tau_{ji}$, which is proportional to the weight of the simplex $\omega_j$. In our analysis we use two different weighting schemes, namely, with a) the bias towards lower order simplices and b) the bias towards higher order simplices.}
    \label{fig:RW_sketch}
\end{figure}
In addition, in many contexts we are required to make quantitative comparisons within and across datasets. 
Persistent homology~\cite{ghrist2008barcodes, petri2013topological} was developed for this purpose of quantifying the shape of data by providing a multiscale description of its topological structure. This is achieved by encoding data as a sequence of simplicial complexes, called a filtration.
Features of the complexes (connected components, one-dimensional cycles, three-dimensional cavities) persist along this sequence of simplicial complexes for different intervals, and their persistence defines how relevant they are for the shape of the dataset. In such a way, it is possible to identify unique meso-scale structures otherwise invisible to classic analytical tools.
This intervals can also be mapped in a persistence diagram~\cite{carlsson_topology_2009} and distances between persistence diagrams can be used. Notably, the persistence diagrams come with various notions of metric: in this work, we will adopt one of the most flexible ones, the (sliced) Wasserstein distance $d^{SWS}$ ~\cite{mileyko2011probability}, which is often used to compare diagrams.

\subsection{Random Walks and symbolic embedding} 
Random walks (RWs) are sequences of locations visited by a walker on a substrate. This substrate can be anything from a metric space to a discrete structure, like a network. Because of this flexibility, RWs are often used as exploration processes for different properties of the substrate. In particular, RWs can be unbiased or biased. 
In the former case, walkers have no preference for the future direction and choose uniformly at random from the possible next locations for a jump, e.g. among the neighbours of the node the walker is sitting on. 
Biased and correlated models were developed to take into account that, in certain situations, some directions are more probable than others and the situations where the sequential step orientations are correlated \cite{Patlak1953}. 

In order to generate the symbolic sequences that will be processed by word2vec in the embedding process, we employed both the unbiased and biased approaches to random walk on the Hasse Diagram (for our purpose, we discard the direction of the inclusions in the DAC when performing a jump). In the unbiased model, the walker starts from any node of the graph and has uniform probability to move to any of the neighbours. This is the RW scheme used for the synthetic simplicial complexes analysis in section \ref{sec:random}.

In the biased model we attach specific weights to nodes of the Hasse diagram and transition probability at each step is proportional to the weight of the destination node (see Figure \ref{fig:RW_sketch}). Specifically, Whereas the node weighting scheme can be in principle arbitrary, we implement two specific weighting schemes used in the analysis: a) with the bias towards lower order simplices (lower order bias) and b) to higher order simplices (higher order bias). The higher order bias scheme is described as follows. Initially all nodes in the Hasse diagram has assigned weight zero and we by start sequentially considering simplices in the data. Each $n$-simplex in the data adds weight 1 to a respective node in the Hasse diagram, the weight $\frac{1}{n+1}$ to $(n-1)$-simplices connected to the given $n$-simplex, the weight $\frac{1}{(n+1)n}$ to $(n-2)$-simplices connected to respective $(n-1)$-simplices and so on. Overall, if the $(n-k)$-simplex $j$ is directly reachable from the considered $n$-simplex, it receives the weight 
$$\omega_j = \frac{1}{(n+1)n\dots (n-k+2)}.$$ 
The lower order bias scheme is analogous to the higher order scheme, except the additional weigths $\overline{\omega}$ are inverse: 
$$\overline{\omega_j} = \frac{1}{\omega_j}.$$
The motivation behind the weighting schemes is to compare two different exploration strategies - the lower order bias tends to explore 0-simplices, whereas the higher order bias pushes the exploration to higher order simplices in order to better capture the higher order structure.

While the biased random walk we implemented is motivated by the combinatorial structure of the inclusions among simplices in the Hasse Diagram itself, any weighting strategy for RW can be chosen according to the needs of the user. 
The length $L$ and the number of walks $n_w$ are important parameters that influence the results of the analysis as well, and usually should be chosen such that  $L n_w \sim |\mathcal{K}|$, where $\mathcal{K}$ is a complex and its cardinality corresponds to the number of simplices in it (or nodes in its Hasse Diagram). 

Notice that in our formulation, RWs are essential to navigate the $k$-simplices of the simplicial complex. As a matter of fact, the sequence of locations visited by a RW can be regarded as a sequence of symbols that encode the local structure of adjacency of the substrate. For example, random walks are at the core of graph embedding techniques, e.g. node2vec \cite{GOYAL201878} and DeepWalk \cite{perozzi2014deepwalk}. These rely on word embeddings developed in the context of natural language processing and unsupervised learning, out of which the most commonly used is word2vec \cite{2013arXiv1301.3781M}.

These models, usually, consist of two-layer neural networks that are trained to reconstruct linguistic contexts of words. Indeed, word2vec takes as its input a large corpus of texts and produces a vector space, typically of several hundred dimensions, with each unique word in the corpus being assigned a corresponding vector in the space. Word vectors are positioned in the vector space such that words that share similar contexts in the dataset are positioned close to one another in the space \cite{goldberg2014word2vec}. 
Word2vec can both predict a word from its context or predict the context from a word. Details on the two main implementations can be found in \cite{Mikolov:2013:DRW:2999792.2999959}. 

DeepWalk leverages random walks to sample node contexts (their quasi-local environment), that are subsequently fed to word2vec in order to calculate the embedding of the nodes.
The DeepWalk method computes a graph embeddings in two steps.
A graph is first sampled with random walks. Then, random walks are treated as sentences in the word2vec approach and fed to its neural network. Node2vec is a modification of DeepWalk. In the node2vec approach, two additional parameters are employed in order to implement an exploration-exploitation trade-off mechanism. One parameter defines how probable it is that the random walk will discover the undiscovered part of the graph, while the other one defines how probable it is that the random walk will go back to the node it visited in the step before. 

\subsection{Simplex2Vec}
The method we propose here, Simplex2Vec, builds on these ideas and it is based on constructing contexts for the nodes of a graph which may then be fed to the word2vec package. More specifically, we use RWs defined on the Hasse diagram associated with a simplicial complex. In such a  way,  by navigating through the $k$-simplices, each walk preserves high-order information on the topological structure of the simplicial complex.  We sample a large set of random walks on the Hasse Diagram, which can be regarded as symbolic sequences or ``words'', and subsequently we embed the contexts created in this manner in a metric space by computing a vector representation (see Figure \ref{fig:Simplex2Vec} for a graphical representation of the workflow). Note that since our RWs are defined on the Hasse diagram, our method simultaneously returns a vector representation for vertices and all of the higher-dimensional simplices (up to the chosen maximal simplicial dimension).
A preliminary implementation of the Simplex2Vec method is available at \url{https://github.com/lordgrilo/doublenegroni}. 

\subsection{Data processing and simplicial construction}
\label{sec:simpconstruction}

\subsubsection{Costa-Farber Model} \label{costa_farber}
To explore the effect of higher order simplices on the embeddings and corresponding partitions, we use the Costa-Farber random simplicial complex model \cite{costa2016random}. 
They proposed a simple construction, based on a flexible model for random simplicial complexes with randomness in all dimensions. 
It starts with a set of $N$ vertices and retains each of them with probability $p_0$ (usually set to 1); then it connects every pair of retained vertices by an edge with probability $p_1$, and then fills in every 3-clique (closed path of edges) with a 2-simplex (a full triangle) in the obtained random graph with probability $p_2$, and so on for constructing simplices of order $k$ with probability $p_k$. At the final step we obtain a random simplicial complex which depends on the set of probability parameters $(p_0, p_1, . . . , p_k)$, $0 \leq p_i \leq 1$ for all $i = 1,\ldots,k$.\\

\subsubsection{Sociopatterns Data}
\label{sec:socio}
We consider a dataset of face-to-face interactions between children and teachers of a primary school (LyonSchool) \cite{gemmetto2014mitigation} and interactions between students of a high school (Thiers13) \cite{Genois2018}, available from \textit{sociopatterns.org}. 
Interactions have a temporal resolution of $20$ s, but we aggregated the data using a temporal window of $\Delta t = 15$ min. Within each time window we define a graph by aggregating the pairwise interactions and computed all the maximal cliques that appear in that time-window. 
These maximal cliques can be considered \textit{facets}. We collect them across all windows and build a simplicial complex by aggregating over all the time windows. We are able to subsequently reconstruct all the lower-order simplices stemming from them, thus recovering the whole structure of the Hasse diagram. 
Note that, for computational reasons in the following analysis, we limit ourselves to consider simplices up to order $3$. While higher-dimensional cliques are not included in the final simplicial complex, their sub-cliques up to size $4$ are considered in the counting.

\subsubsection{Functional Connectivity}
\label{sec:fmri}
Furthermore, we consider a dataset composed of group average fMRI time series correlations ($n=819$). These correlations are understood to characterize relationships among brain regions, i.e., their degree of "functional connectivity" \cite{SMITH2013144,BILLINGS2017344,Ganiesse2019}. 
Data were acquired from the Human Connectome Project \cite{VANESSEN20122222}. 
Specifically, we used an adjacency matrix of group-level correlations among 200 brain regions as input to the Hasse graph. The 200 regions are a projection from the raw fMRI voxel space onto 200 independent components (ICA200). The native connectome matrices (netmats) arrive as $z$-statistics of Pearson correlation $r$-values. Data were therefore preprocessed to normalize inputs to the range $[-1,1]$ by calculating the inverse Fischer transform, with all constant terms set to $0.18$. We then calculated absolute values on the adjacency matrix because we were interested in describing $k$-dimensional coordinations (correlation with a 180 degree phase shift) among ICA components rather than distinguishing between strongly positive versus strongly negative correlations. This is because coordinated activation and suppression of brain hemodynamics develops unique and important routes for information flow with respect to a given brain state. Finally, the adjacency matrix was masked with a threshold of the upper $5\%$ of correlation values. Facets were constructed from the maximal cliques of the thresholded graphs, and included all sub-cliques. Equal weights were given to the random-walk transition probabilities between facets connected by an edge in the Hasse graph. 
\subsection{Clustering}
When considering a graph structure, clustering of the nodes is usually carried out by directly probing the topological structure of the graph itself. This task goes under the name of community detection, and many algorithms have been developed for this purpose \cite{newman2006modularity,rosvall2008maps,blondel2008fast,traag2019louvain}. In the case of simplicial complexes however it is less clear how one might want to re-define classical community detection algorithms. 
Against this background, embeddings are very helpful: indeed, one of the main purposes of embeddings is to represent data in a non-trivial metric space, where it becomes easier for clustering algorithms to partition data into sensible groups. 
In order to test the quality of our embedding procedure, we run the widespread \textit{hierarchical clustering} algorithm~\cite{friedman2001elements} on its representations of nodes. We then proceed to test the quality of the obtained results against the true labels of groups of nodes, by means of  the normalized mutual information ($NMI$) metric~\cite{cover2012elements}. Even though more sophisticated clustering algorithms have been devised during the years, the good results that can be obtained already with this relatively simple procedure show the validity of our embedding approach. 
\section{Results}
\subsection{Evolution of embedding structure}
\label{sec:random}
\begin{figure}[t]
    \centering
    \includegraphics[width=0.85\columnwidth]{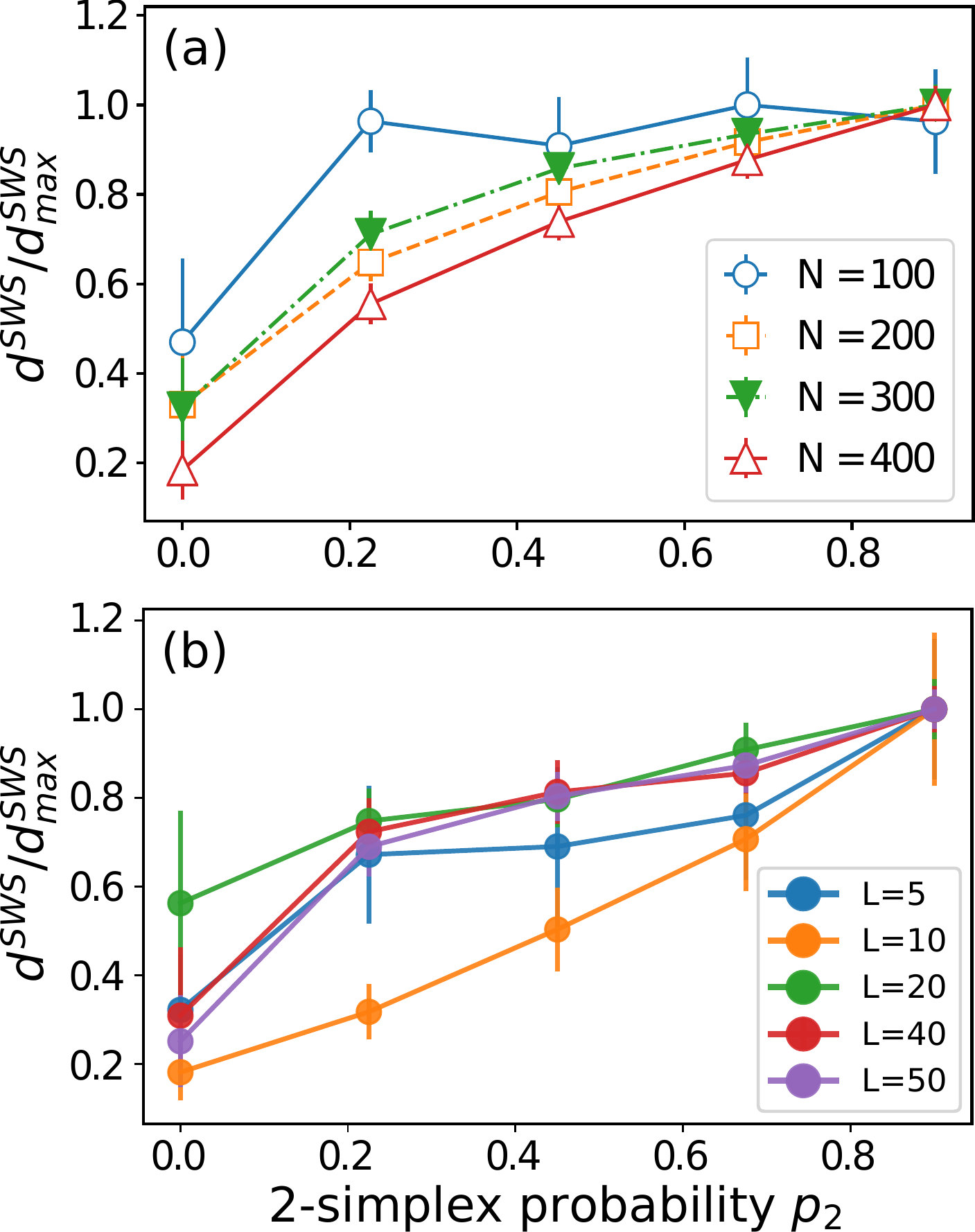}
    \caption{\textbf{Effect of higher order interactions.} (a) Sliced Wasserstein distance $d^{SWS}_1(X[p_1,0], X[p_1,p_2])$ between one-dimensional persistent homology group $H_1$ of the embedding corresponding to a pure network ($p_2=0$) and the embedding obtained for simplicial complexes as a function of $p_2$ (walks length $L=30$) and the number of nodes $N$ of the network. (b) $d^{SWS}_1$ distance for the embeddings of Costa-Farber simplicial complexes as a function of $p_2$ and different walk lengths. We find that for walk lengths $L \ll N$ the embeddings are stable, and display smaller changes as compared with the effect of $p_2$ (number of nodes $N=200$). For visualization purposes, for fixed $N$ in (a) or $L$ in (b), we normalize $d^{SWS}$ by the maximum distance value $d^{SWS}_{max}$ observed.}
    \label{fig:embedding_stability}
\end{figure}

The first step toward understanding the role of higher order interactions in shaping the resulting embedding is quantifying the difference between the embeddings obtained with and without the contribution of higher order interactions. 

To investigate this, we construct a series of simplicial complexes using the Costa-Farber random simplicial complex model (described in the Methods, section \ref{costa_farber}). In this case, the model has only two parameters: the probability $p_1$ of two nodes being connected (exactly like in the standard Erdos-Renyi model), and the probability $p_2$ that a 3-clique in the underlying 1-skeleton will be filled by a 2-simplex.
In particular, when $p_2=0$ the simplicial complex is one-dimensional and corresponds therefore to a standard graph. 
Hence, for fixed number of nodes $N$ and edge probability $p_1$, we can investigate the effect of introducing higher order simplices by gradually increasing $p_2$.

We quantify the effects of increasing $p_2$ by computing the one-dimensional persistence homology of the embeddings (using the standard Rips-Vietoris filtration) and then calculating the sliced Wassersstein distance between the resulting persistence diagrams (Figure \ref{fig:embedding_stability}a). 
We find large changes in the embedding topology already for small $p_2$ values; in particular, the variability of topology across different realizations of the Costa-Farber model for the same $p_2$ value is much smaller than change induced by increasing $p_2$. 
We show the same general behaviour for different number of vertices $N$ (Figure \ref{fig:embedding_stability}a) and for the walk lengths $L$, which is one of the main parameter of the embedding construction (Figure \ref{fig:embedding_stability}b).  

\subsection{Effects of higher-order interactions on clustering stability}
\begin{figure*}
    \centering
    \includegraphics[width=.75\textwidth]{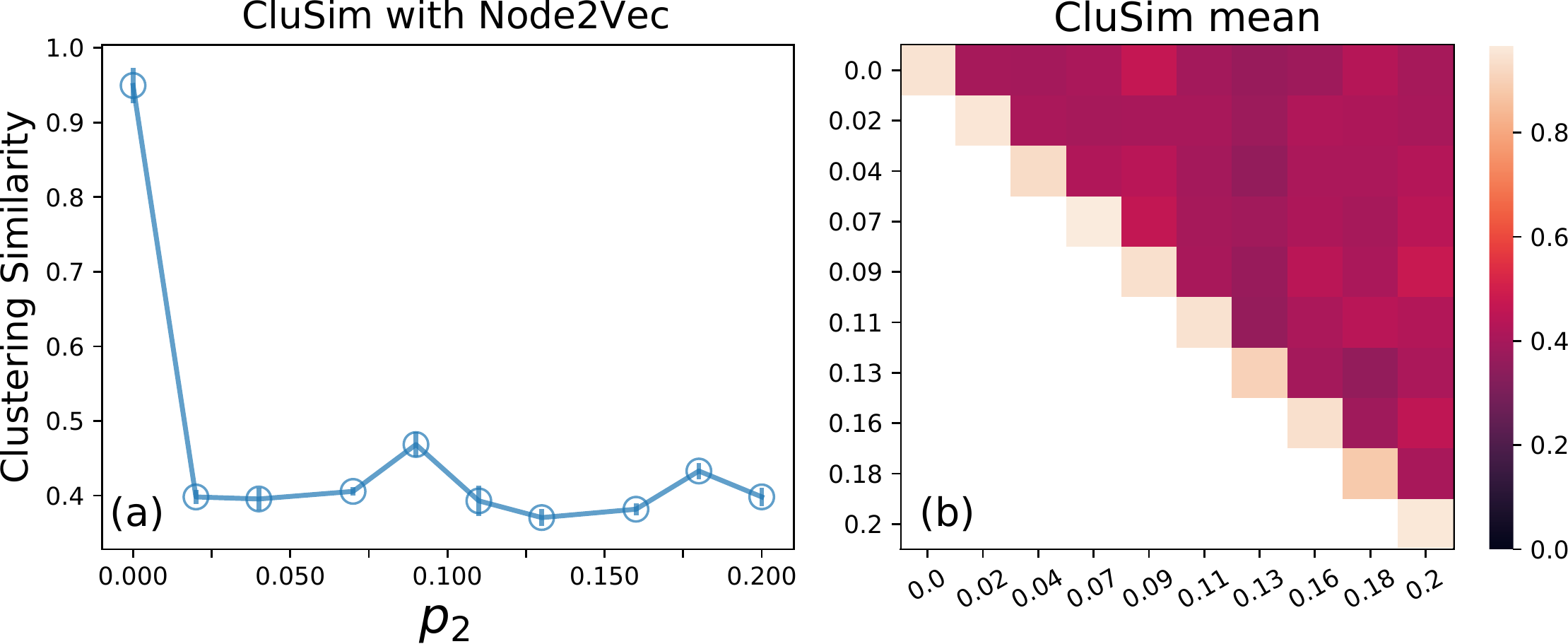}
    \caption{\textbf{Higher-order interactions disrupt cluster structures.} (a) Clustering similarity between the cluster partitions obtained at $p_2=0$ and those obtained for positive $p_2$ ($N=200$). Interestingly, even with a small increase in  $p_2$, we clearly find that the partitions obtained are very dissimilar from the one obtained considering only the edge structure. Analougous results are obtained when comparing partition across different values of $p_2$ (b).}
    \label{fig:community_stability};
\end{figure*}

While we showed that the overall topology of the embedding is stable over realization of the same random model, this does not necessarily imply that the local structure of the embedding is stable too. 
This is important to understand however because the local relations between the embedded points are relevant for the resulting clustering partition. We check this as follows: we construct a graph, that is, an instance $X$ of the Costa-Farber model without triangles ($p_2=0$); then starting from the same graph, we add triangles with different $p_2$ and compute the corresponding embeddings. 
We find that already when small amounts of higher-order interactions are present, the partitions are very dissimilar from the one obtained considering only the edge structure (Figure \ref{fig:community_stability}a). The same is true when we compare across different values of $p_2$. 
We find in fact that the partition structure shows consistently a low similarity across different triangle densities (i.e. average similarity $\sim0.4$, Figure \ref{fig:community_stability}(b)). 
Note also that different realizations with the same triangle density (the diagonal in Figure \ref{fig:community_stability}(b)) result in partitions that are much closer than those obtained for different triangle densities, once more highlighting the critical role of higher order interactions. 

\subsection{Simplicial communities in social systems}
We test now whether using the information contained in high-order interactions is relevant to obtain information about the community structure of a simplicial system. 
We investigate this by considering a datasets of temporal face-to-face interactions collected by the Sociopatterns collaboration: data from contacts between students in an elementary school and in a high school \cite{10.1371/journal.pone.0136497,Gemmetto2014}. 

Following Iacopini et al. \cite{iacopini2019simplicial} (see section \ref{sec:socio}), we aggregate interactions over small non-overlapping time windows (15 minutes) in sub-networks.
We then consider each fully connected sub-graph of interactions created within a window as a simplex and construct the simplicial complex aggregating all interactions across all windows.

\begin{figure*}
    \centering
    \includegraphics[width=0.75\textwidth]{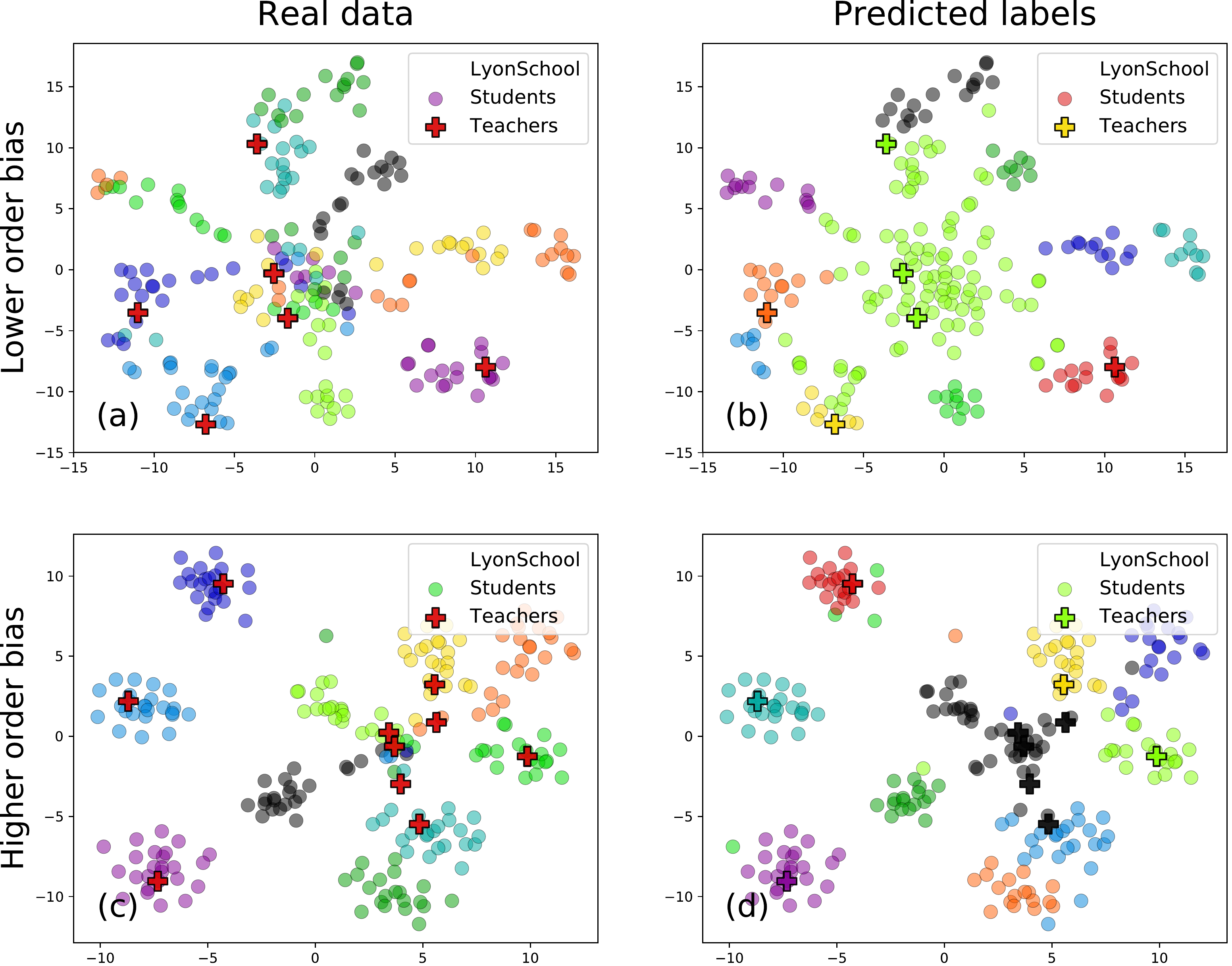}
    \caption{\textbf{Simplicial Communities for face-to-face interaction data in a primary school.} We present the Simplex2Vec embedding of group gatherings of students of a primary school (LyonSchool). The real labelling of the node embedding is represented in (a) and (c), where the colors represent attribution to the same class and crosses ($+$) denote teachers. Random walk bias towards lower order simplices (b) produces less distinguishable embedding in comparison with the higher order bias (d). Higher order bias also permits to glance over the teachers attribution to classes. The coloring order between real data and predicted labels does not match and presented only for the sake of partition distinguishability.}
    \label{fig:sociopatterns_example1}
\end{figure*}

\begin{figure*}
    \centering
    \includegraphics[width=0.75\textwidth]{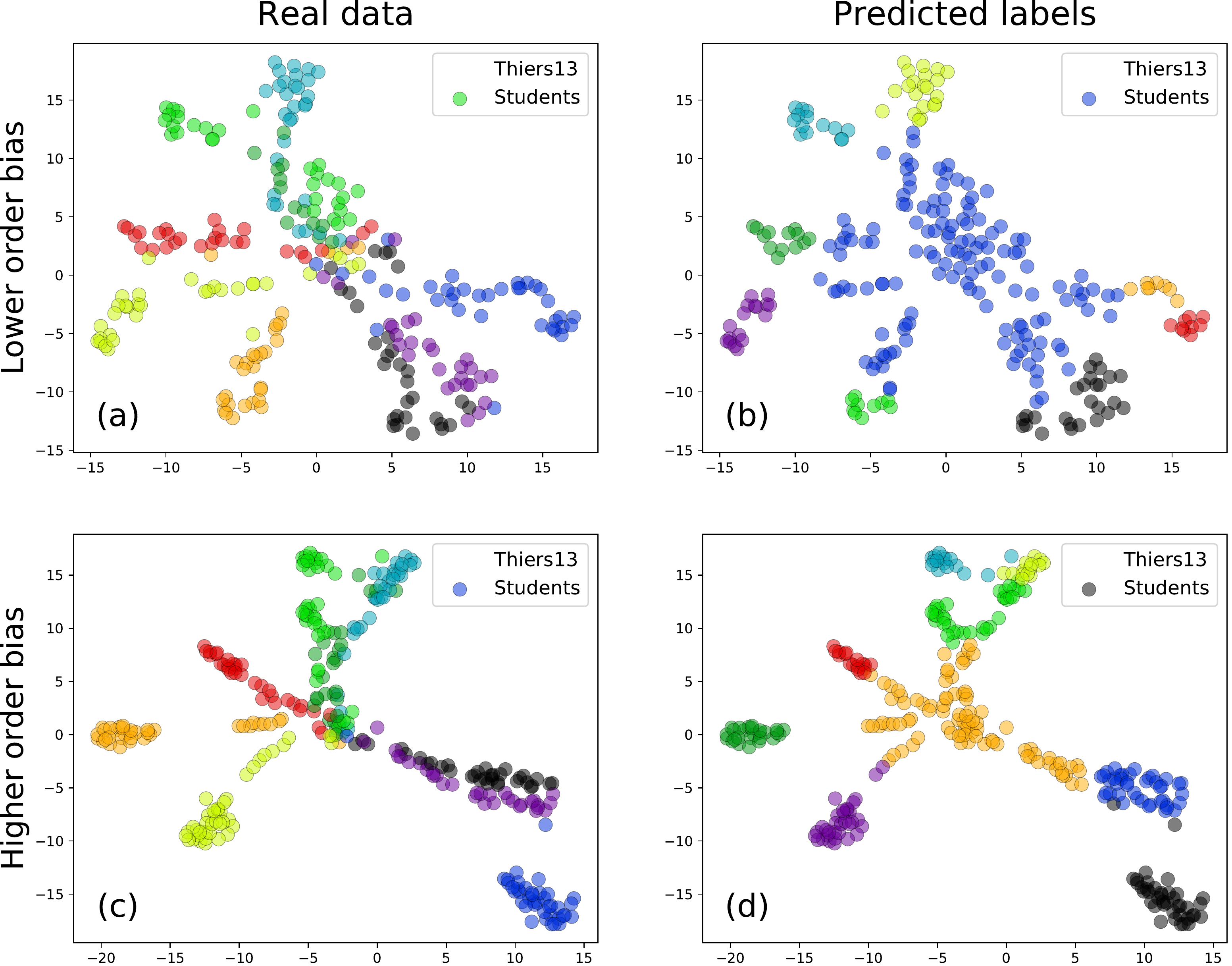}
    \caption{\textbf{Simplicial communities for face-to-face interaction data in a high school.} We present the Simplex2Vec embedding of group gatherings of students of a high school (Thiers13). The real labelling of the node embedding is represented in (a) and (c), where the colors represent attribution to the same class. This dataset does not have separate labels for teachers. Random walk bias towards lower order simplices (b) produces less distinguishable embedding in comparison with the higher order bias (d). The coloring order between real data and predicted labels does not match and presented only for the sake of partition distinguishability.}
    \label{fig:sociopatterns_example2}
\end{figure*}

We compute the Simplex2Vec embedding for each dataset and detect clusters in the embeddings using standard clustering techniques. 
In Figures \ref{fig:sociopatterns_example1} and \ref{fig:sociopatterns_example2} we display the results of the clustering detection on the embeddings built using simplices up to dimension $k=4$ for sociopattern datasets and compare them with data on which class the students belong to. 
Both panels contains results for both the weighting schemes,  biased toward lower order simplices (top row), and biased toward higher order simplices (bottom row). We find a better NMI ($\sim 0.5$ for the second case highlighting again the role of higher order simplices in detecting the large and meso-scale structure of the system. 

While these results are already encouraging, it is interesting to quantify how much information is encoded in higher-order interactions. 
This can easily be capping the maximal simplex dimension that we allow in the Hasse Diagram used to construct the embedding. 
In other words, we repeat the construction of the embeddings but considering only simplices up 1-simplices (edges), then up to 2-simplices (triangles), then up to 3-simplices (tetrahedra) and so on. 

\begin{figure*}
    \centering
    \includegraphics[width=0.75\textwidth]{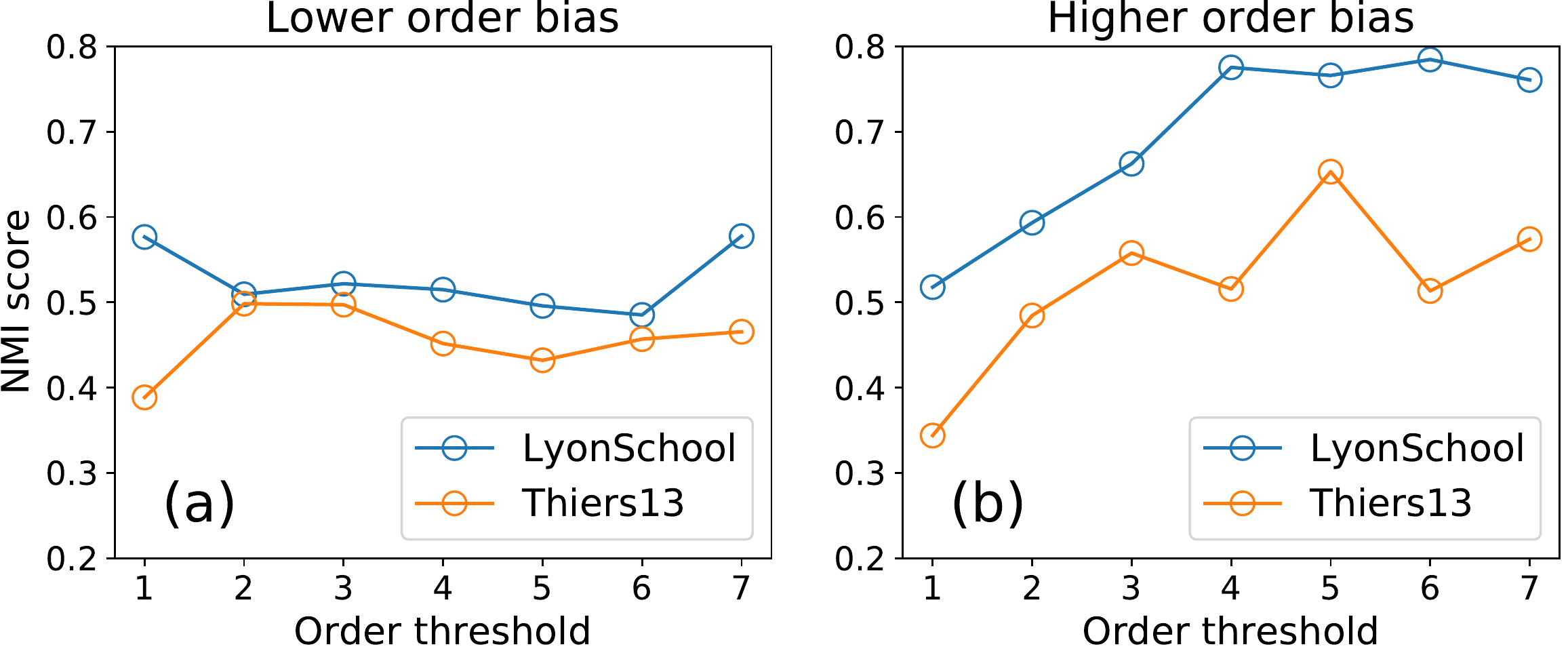}
    \caption{\textbf{Similarity between simplicial embedding partitions and ground data grows with maximal simplicial dimension.}. We plot the NMI values for the partition (obtained via agglomerative clustering) from the Simplex2Vec embeddings using random walks that are biased towards (a) lower order simplices and (b) higher order simplices for two datasets of face-to-face interactions. 
    We clearly find that biasing towards higher order simplices results in an improved NMI between the detected clusters and the class ground truth.}
    \label{fig:sociopatterns_nmi}
\end{figure*}

This provides us with a natural complexity ladder for the simplicial embeddings. 
For each maximal simplicial dimension, we repeat the whole construction and compute the corresponding clusters. 
As expected, we find that, when we bias toward lower-order simplices, increasing the order of the Hasse Diagram has little to no effect on the resulting NMI. In contrast, when using considering the bias toward higher order simplices, we find a clear increase in the resulting NMI (Figure \ref{fig:sociopatterns_nmi}).

\subsection{Communities of fMRI functional connectivity}
\begin{figure*}
    \centering
    \includegraphics[width=.9\textwidth]{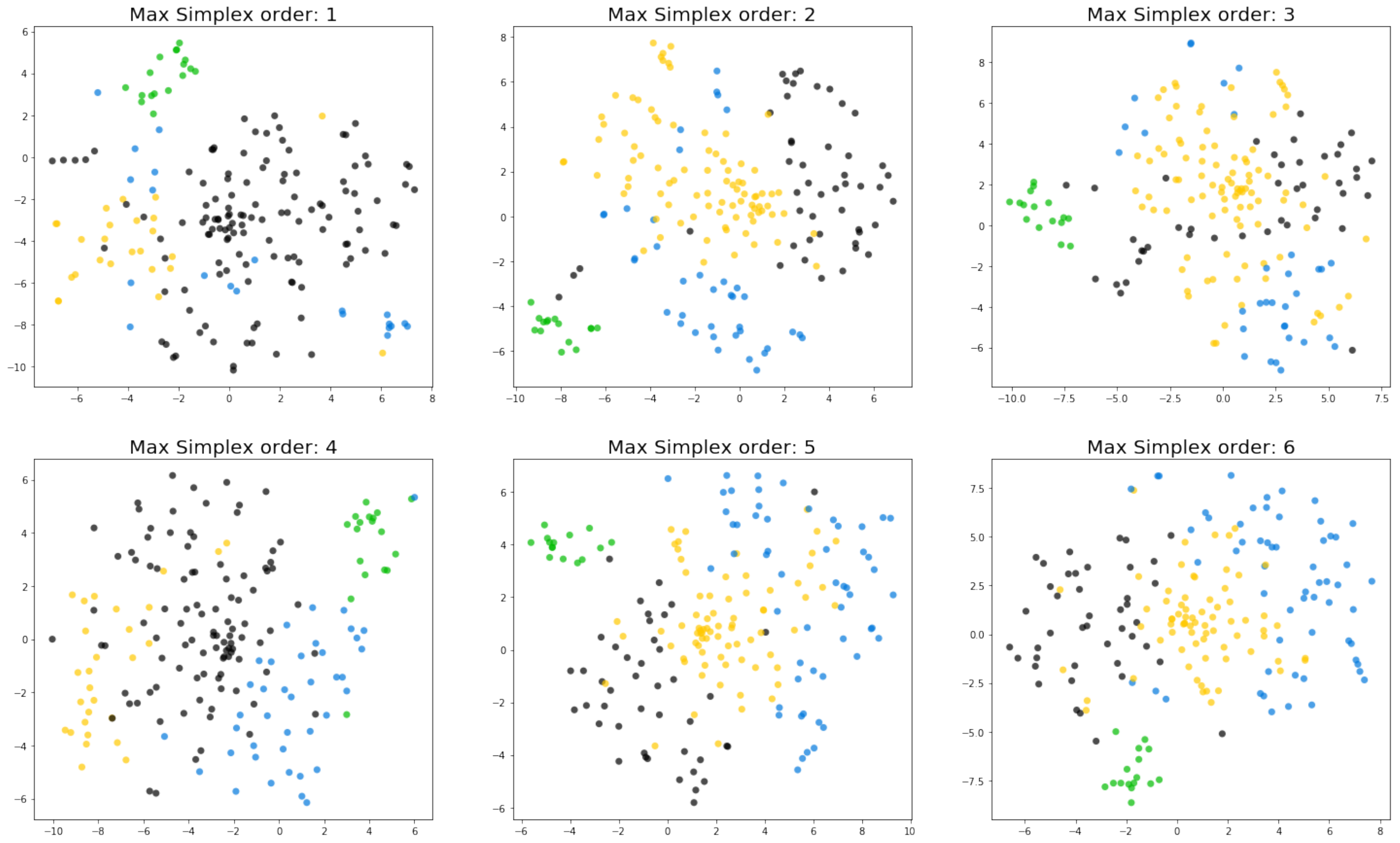}
    \caption{\textbf{Functional Connectivity Embeddings.} We show the results of the Simplex2Vec construction for increasing maximum simplicial dimension. We find a large variability across brain regions in the embedding and clustering (colors) results. }
    \label{fig:fmri}
\end{figure*}

We computed Simplex2Vec embeddings for the average functional connectivity matrix described in section \ref{sec:fmri}
the embedding using different maximal simplicial dimensions in the Hasse Diagrams. We then perform clustering over the different embeddings and show the results in Figure \ref{fig:fmri} (only points corresponding to brain regions are showed). 
We find a wide variability in the results of both the embedding and clustering. We use this variability to ask which regions display a conserved community participation structure across different maximal simplicial orders. 
For each region $i$ and maximum simplicial order $d$, we compute a community participation vector $\mathbf{c}_i(d)$. We then compute the cosine similarity between the $\mathbf{c}_i(1)$ corresponding to only considering edges, and all the higher order forms, $s_i =\langle \text{cos}(\mathbf{c}_i(1), \mathbf{c}_i(d))\rangle_d$ averaged over all $d \in (2,6)$. 

We use $s_i$ as a coarse indicator of which regions have integration patterns that are prone to be affected by higher order interactions (low $s_i$), and regions that are less prone to be affected by higher order interactions. Regions strongly affected across repetitions include the reticular formation, the secondary visual cortex (including early areas of the dorsal and ventral streams), the conjunction of the secondary somatosensory cortex and thalamic nucleai, and single components that highlighted distributed nucleai in the brain stem. A very large plurality of regions in the cerebellum were stable across maximal interaction orders, as were focal brain stem nuclei, large patches of the bilateral dorso-ventral thalamus, and the caudate nucleus.

\section{Discussion}
Embeddings are a very general tool to study the structure and --potentially-- dynamics of interacting systems, thanks to their capacity to represent discrete systems in continuous spaces. 
This motivated intense research to develop informative and scalable algorithms able to construct such embeddings for unstructured and --more recently-- network data.
However, with increasing frequency and across a range of disciplines, datasets are produced in which interactions are not well described by pairs of agents, but rather involve a larger and at times heterogeneous number of agents, with applications ranging from structure and prediction in temporal networks \cite{petri_icial_2018,benson2018simplicial} to contagion dynamics \cite{iacopini2019simplicial}. 
Simplicial complexes represent a promising way for representing such interactions, but the identification of communities remains an important challenge. The current work takes a step towards filling this gap, by providing a solution based on the developments in symbolic embeddings. 

We showed in simple case studies that higher order simplices can alter strongly the community structure of an interacting system and should therefore be considered (or at least their contribution tested). We provide evidence from a social system that simplicial contributions are important to improve the correspondence of the detected clusters to the ground truth. 
Finally, using a parsimonious example from network neuroscience, we we showed how to assess which nodes in a system are more (or less) prone to be affected by higher order interactions. 

Naturally, due to time constraints, this work has many limitations, which we plan to address in the nearest future. 
In particular, interesting avenues for future work include:
\begin{enumerate}
    \item extending the Simplex2Vec analysis to other dynamical systems, where the procedure described in section \ref{sec:socio} is relevant to extract group interactions over small temporal scales;
    \item testing the capacity of Simplex2Vec to predict missing interactions in arbitrary dimension: since Simplex2Vec simultaneously embeds all interactions, it is possible to generate predictions for any group size;
    \item extending the community analysis to higher order simplices: this has already been done to some degree by considering edge-communities (in neuroscience for example, see \cite{battaglia2017functional}), but Simplex2Vec provides a general framework for this;
    \item exploring a wider range of RW navigation biasing schemes (for example, maximally entropic RW\cite{sinatra2011maximal}) and include data-driven weights on simplices; 
    \item and, finally, relating the results of unconstrained RW with those constrained to moving using the combinatorial Laplacian \cite{muhammad2006control,schaub2018random} (walks constrained to simplices in dimension $k\pm1$).
\end{enumerate}

\section{Acknowledgments}
This work was produced by the \textit{DoubleNegroni} group at Complexity72h Workshop, held at IMT School in Lucca, 17-21 June 2019. Website: \textit{complexity72h.weebly.com}

Data were provided [in part] by the Human Connectome Project, WU-Minn Consortium (Principal Investigators: David Van Essen and Kamil Ugurbil; 1U54MH091657) funded by the 16 NIH Institutes and Centers that support the NIH Blueprint for Neuroscience Research; and by the McDonnell Center for Systems Neuroscience at Washington University.
A.M. acknowledges support of the Fonds de la Recherche Scientifique-FNRS under Grant n. 33722509 (CollectiveFootprints) and the grant 19-01-00682 of the Russian Foundation for Basic Research.
G.P. and JCW. B.acknowledges partial support from Compagnia San Paolo (ADnD grant).
F.M. acknowledges the support of \textit{Istituto Nazionele di Fisica Nucleare} (INFN), under the \textit{BIOPHYS} experiment grant.
\FloatBarrier
\bibliographystyle{unsrt}

\end{document}